# The Miniature X-ray Solar Spectrometer (MinXSS) CubeSats: spectrometer characterization techniques, spectrometer capabilities, and solar science objectives.

Christopher Samuel Moore*[abc], Thomas N. Woods[b], Amir Caspi[d], James Paul Mason[b,e]
[a]Astrophysical & Planetary Sciences Department, University of Colorado, Boulder, CO 80309, USA
[b]Laboratory for Atmospheric & Space Physics, University of Colorado, Boulder, CO 80303, USA
[c]Center for Astrophysics & Space Astronomy, University of Colorado, Boulder, CO 80309, USA
[d]Southwest Research Institute, Boulder, CO 80302, USA
[a]Aerospace Engineering Sciences Department, University of Colorado, Boulder, CO 80309, USA
*Email: Christopher.Moore-1@Colorado.edu

## ABSTRACT

The Miniature X-ray Solar Spectrometer (MinXSS) are twin 3U CubeSats. The first of the twin CubeSats (MinXSS-1) launched in December 2015 to the International Space Station for deployment in mid-2016. Both MinXSS CubeSats utilize a commercial off the shelf (COTS) X-ray spectrometer from Amptek to measure the solar irradiance from 0.5 to 30 keV with a nominal 0.15 keV FWHM spectral resolution at 5.9 keV, and a LASP-developed X-ray broadband photometer with similar spectral sensitivity. MinXSS design and development has involved over 40 graduate students supervised by professors and professionals at the University of Colorado at Boulder. The majority of previous solar soft X-ray measurements have been either at high spectral resolution with a narrow bandpass or spectrally integrating (broadband) photometers. MinXSS will conduct unique soft X-ray measurements with moderate spectral resolution over a relatively large energy range to study solar active region evolution, solar flares, and the effects of solar soft X-ray emission on Earth's ionosphere. This paper focuses on the X-ray spectrometer instrument characterization techniques involving radioactive X-ray sources and the National Institute for Standards and Technology (NIST) Synchrotron Ultraviolet Radiation Facility (SURF). Spectrometer spectral response, spectral resolution, response linearity are discussed as well as future solar science objectives.

## 1. SOLAR SCIENCE WITH THE MINIATURE X-RAY SOLAR SPECTROMETER (MINXSS) CUBESATS

The solar corona is the interface between the solar surface (photosphere) and lower atmosphere (chromosphere), and the solar wind (interplanetary space), starting at roughly 2,000 km above the solar photosphere and extending to a few solar radii[1]. The conditions in the corona are drastically different than the environment in the photosphere. The temperature in the corona reaches above $10^6$ K in contrast to a photosphere temperature around 5,700 K. Furthermore, the density varies from as low as ~$10^8$ particles per $cm^3$ (e.g., in coronal holes) to as high as $10^{11}$ $cm^{-3}$ (e.g., in flaring loops), as compared to a photospheric average density of ~$10^{17}$ $cm^{-3}$. Additionally, the plasma beta ($\beta = P_{gas}/P_{magnetic}$), the ratio of gas pressure to magnetic pressure, that dictates which force dominates the dynamics, transitions from high (gas force dominated) in the photosphere, to low (magnetic force dominated) in the corona. This demonstrates the drastic changes in atmospheric conditions encountered in the corona even when not considering eruptive transient processes.

The high temperature, low density, and dynamical conditions result in highly ionized plasma, emitting primarily UV and X-rays that is mostly optically thin to soft X-rays (besides specific scattering events) and is confined by magnetic fields. The soft X-ray emission includes both spectral lines (bound-bound emission) from the various ion species in the corona, and continuum emission including free-free, electron-ion bremsstrahlung and free-bound, radiative recombination emission. The specific spectral emissions depend uniquely on the coronal conditions. Thus, Soft X-rays present a rich diagnostic data set, and understanding the solar coronal soft X-ray flux yields insight on plasma densities, temperatures, chemical abundances, magnetic field structure and velocity flows. Spectrally, temporally and spatially resolved measurements are necessary to answer fundamental questions such as:

1. What is the solar soft X-ray spectral energy distribution and how does it vary over a solar cycle?
2. How do solar active region physical parameters evolve over time as inferred from soft X-rays?
    a. Temperature distribution of emitting plasma (differential emission measure)
    b. chemical abundance
    c. magnetic field morphology
3. What is the solar flare soft X-ray energy distribution and how does it vary with flare size (in addition to the parameters listed in 2)?
4. What are the primary heating mechanisms contributing to the observed soft X-ray flux?

While there have been many space-, rocket- and balloon-borne missions to measure X-rays to address these *very broad* questions, no single current (or planned in the foreseeable future) mission will meet all the performance requirements to address all the science objectives listed above in significant detail, but a series of well-designed instruments conducting observations from different observatories could. Science oriented CubeSats like the Miniature X-ray Solar Spectrometer (MinXSS)[2] can fill a niche with specific observations to complement those conducted by the larger missions of Hinode[3], the Reuven Ramaty High Energy Spectroscopic Imager (RHESSI)[4], the Solar Dynamics Observatory (SDO)[5], the Geostationary Operational Environmental Satellites (GOES) X-ray Sensor (XRS)[6], the Interface Region Imaging Spectrometer (IRIS)[7] and other large observatories.

In this paper we discuss the first of the two MinXSS CubeSats (from here-on called MinXSS-1) mission objectives, characterization of the X-ray spectrometer on MinXSS-1 and discussion of the improvements possible with the second CubeSat, MinXSS-2. We only focus on basic properties of the spectrometer for this paper and leave the specific details and other instrument characterization for a future paper.

## 1.1 New Spectrally Resolved Soft X-ray Measurements

Figure 1 is a non-exhaustive list of EUV and X-ray space instruments that have conducted spectral measurements between 0.06 and 120 keV (updated from Mason et al. 2016[2]). The majority of these measurements were of either high spectral resolution ($\Delta E < 0.1$ keV FWHM) over a narrow bandpass (e.g. Bragg crystal spectrometers) or low/no spectral resolution over a fairly large bandpass (e.g. integrating photometers). The main MinXSS science performance goal is to conduct accurate and precise spectrally resolved soft X-ray measurements at moderate spectral resolution ($\Delta E \sim 0.15$ keV FWHM at 5.9 keV) over a fairly broad bandpass (nominal sensitivity from 0.5 to 30 keV). An important use of these measurements will be as input for Earth ionospheric (E region) models to better understand the space weather impacts at Earth and on our space-based technology. In addition, MinXSS data will be analyzed to begin addressing the solar science questions stated earlier in the text. MinXSS observations will be extremely valuable, since there are no current observatories conducting spectrally resolved measurements in the 0.5 – 6 keV bandpass. RHESSI makes routine observations down to 3 keV, but with limited performance below 6 keV, with a nominal spectral resolution of ~1 keV.

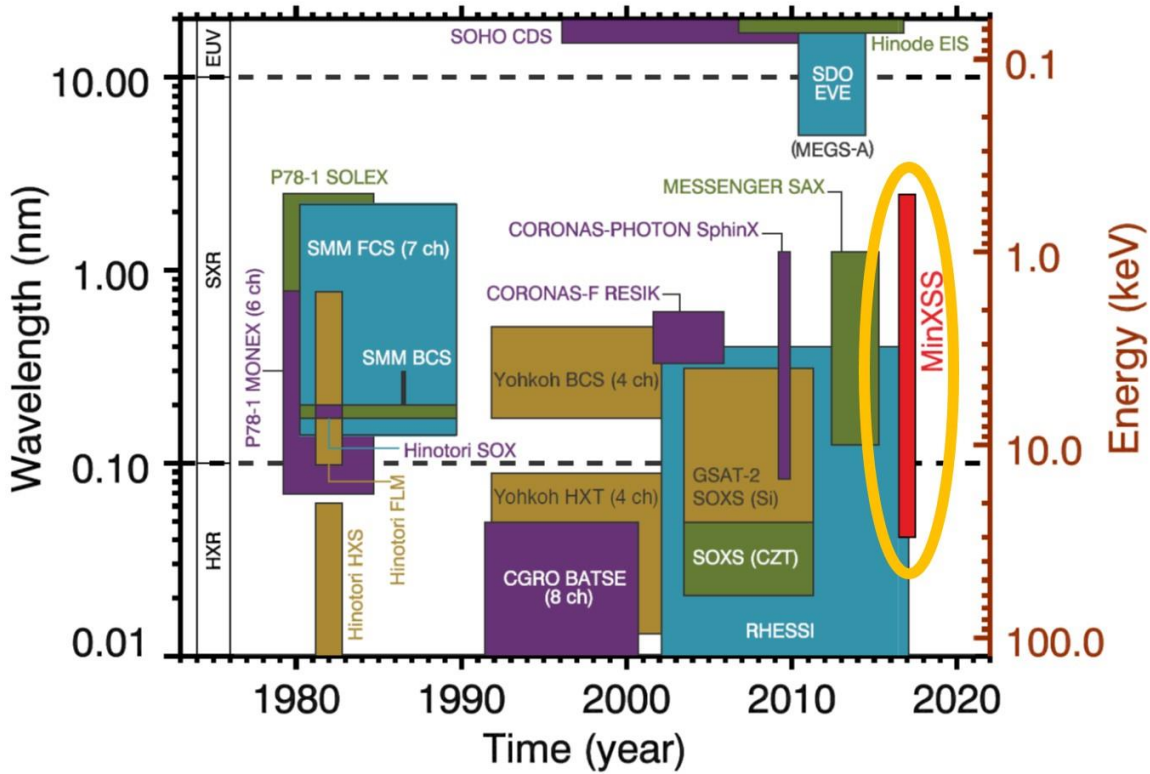

Figure 1. A non-exhaustive list of solar EUV and X-ray space borne instruments. The majority of soft X-ray missions conducted either high spectral resolution (< 0.1 keV) measurements over a narrow bandpass, had fairly low spectral resolution (> 1.0 keV) over a large spectral bandpass, or were integrating photometers. The MinXSS CubeSats will combine moderate spectral resolution (~0.15 keV FWHM at 5.9 keV) and a fairly large spectral bandpass (0.5 – 30 keV).

## 2. THE MINXSS CUBESAT MISSIONS

Figure 2 shows one of the MinXSS[2] twin science-oriented 3U (1U = 10 cm cube) CubeSats that are funded mostly by NASA, but also has had support from NSF, the University of Colorado Boulder (CU) Aerospace Engineering Sciences (AES) Department, and the Laboratory for Atmospheric and Space Physics (LASP). The MinXSS project started as a graduate student course in the CU AES department and has had over 40 students contribute over the project lifetime. Additionally, MinXSS has had substantial input and support from professors, professional staff and postdocs. MinXSS-1 launched to the International Space Station (ISS) as part of the Orbital ATK OA-4 Atlas V resupply mission on December 6, 2015 and was deployed from the ISS on May 16, 2016. MinXSS-1 has been commissioned and, as of June 9, 2016, is in normal science operations mode. The nominal mission operation length for MinXSS-1 is 6 – 12 months depending on solar activity. High solar activity (many flares) causes expansion of Earth's atmosphere, increasing orbital drag and increasing the orbital decay rate because the MinXSS CubeSats do not have any propulsion to change their orbital altitude. This sets the lower bound on the MinXSS-1 mission length (6 months).

In addition to being a solar X-ray science mission, the MinXSS CubeSats are the first technology demonstration flight for the Blue Canyon Technologies 0.5U attitude and determination control system, XACT. MinXSS also consists of many other subsystems including an Electrical Power Systems (EPS) assembly, Command and Data Handling (CDH), and UHF communication with a tape measure radio antenna. Detailed descriptions of all these components are given in Mason et al. 2016[2]. There are three science instruments on MinXSS. The first is the Sun Positioning Sensor (SPS), which consists of a quad Si-photodiode system behind a ND7 filter for visible light position information on the Sun. This is not strictly a science instrument, but provides ancillary information (solar pointing) for science processing. The second instrument is the X-ray photometer called (XP) consisting of a Si-photodiode behind a beryllium (Be) window to provide a spectrally integrated X-ray flux reference measurement. The final, but most scientifically relevant instrument, is the X-ray spectrometer, which is a commercially purchased silicon drift diode (SDD) called X123 from Amptek[8]. This paper is primarily focused on the basic characterization and core capabilities of the X-ray spectrometer, X123, on MinXSS-1. A future paper will describe both MinXSS-1 and MinXSS-2 full suite of science instrument capabilities.

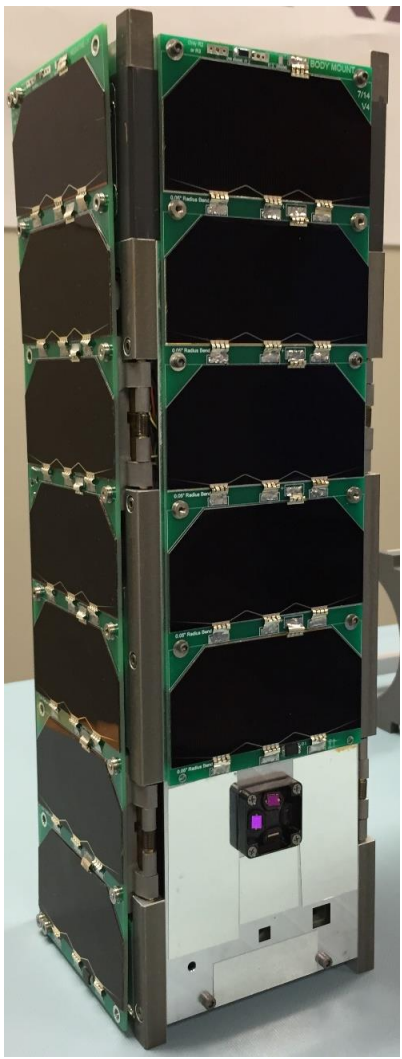

Figure 2. Photograph of the MinXSS-1 CubeSat. MinXSS-1 and -2 are twin 3U science-oriented University of Colorado Boulder CubeSats. MinXSS-1 was launched on the Orbital ATK OA-4 Atlas V resupply mission to the International Space Station (ISS) on December 6, 2015 and deployed from the ISS on May 16, 2016 to a ~400 km Low Earth Orbit. MinXSS-1 has finished commissioning and is currently in normal science operations mode for an estimated 6 – 12 months (depending on Earth atmospheric conditions influencing the deorbit rate). MinXSS-2 is anticipated to launch in the near future for a nominal 4–5 year science mission from a sun-synchronous orbit.

## 3. THE MINXSS X-RAY SPECTROMETER – X123

The commercially purchased Amptek SDD X-ray spectrometer, X123, is the main work-horse for MinXSS science investigations. X123 is an intrinsic energy dispersive detector that constructs spectra from histograms of pulse height analyzed photon events. An incident photon absorbed by the SDD bulk silicon active area generates electron-hole (e-h) pairs and the number of e-h pairs is proportional to the energy of the incident photon (nominal e-h generation rate is 1 e-h pair per 3.65 keV energy deposition in silicon). The electrons are drifted to a central readout anode by a gradient in the applied bias voltage on concentric annular cathodes. The charge collected on the readout anode is converted to a voltage and amplified by a field effect transistor (FET) resulting in a voltage step increase for each photon X-ray event. Each event passes through two parallel digital shaping chains: a fast channel (~100 ns peaking time) and a slow channel (customizable peaking time). The slow channel uses a trapezoidal shaper (set to 4.8 µs peaking time for MinXSS-1) for improved spectroscopy, while the fast channel uses a triangular shaper (with 100 ns peaking time on MinXSS-1) for automatic rejection of overlapping photon events during the slow channel shaping period ("pulse pileup"). The multichannel analyzer probes the fast channel to discern if each event is a unique photon interaction detection by rejecting slow channel events where multiple fast channel events were triggered. Each valid slow-channel photon event is analyzed by the digital pulse processor electronics (DPP) to determine the incident photon energy from the shaped pulse height. The histogram of the valid slow-channel events in an integration period is the resultant spectrum.

Figure 3, courtesy of Amptek, shows the relatively small package size of the X123 system and the basic layout of the detector. X123 consists of a SDD behind a Be window, with a Peltier (thermoelectric) cooler to keep the operational temperature of X123 near 224 K (~-50 °C) and a preamplifier (the FET). The small packaging and relatively low power consumption (~ 5 W max and nominally 2.5 W during normal operations) make X123 optimal for small satellites. We have characterized the performance of our X123s for MinXSS using the radioactive line sources $^{55}$Fe and $^{241}$Am, and continuum photon synchrotron radiation from the National Institute for Standards and Technology (NIST) Synchrotron Ultraviolet Radiation Facility (SURF) in Gaithersburg, MD. Using these light sources, we have determined our nominal detector energy gain (keV/bin) and offset, nominal spectral resolution at ~5.9 keV (for a 4.8 µs peaking time), spectral response, and linearity of response to an input photon flux.

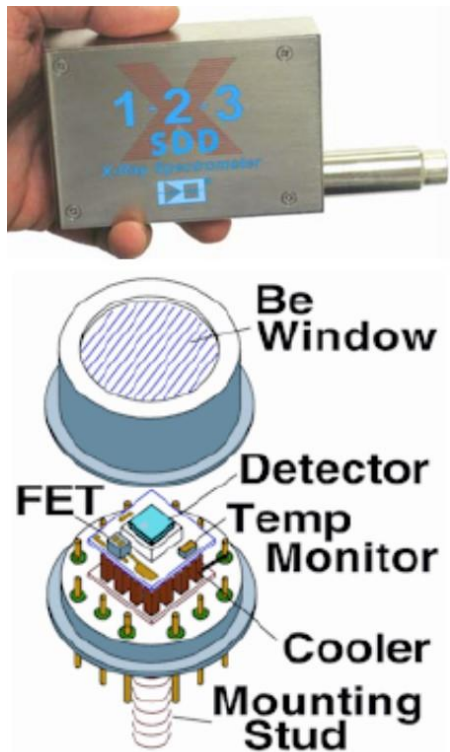

Figure 3. Basic layout of the X123 detector from the Amptek website. The MinXSS X-ray spectrometer is a commercial off the shelf (COTS) X123 silicon Drift Diode (SDD) from Amptek (http://amptek.com/). The detector package comes in various sizes and each can fit in the palm of an adult's hand, making X123 perfect for a CubeSat. The X123 consists of a beryllium (Be) window to attenuate visible light and transmit X-rays, the SDD detector, thermoelectric cooler (Peltier cooling) and a Field Effect Transistor (FET) pre-amplifier all in the detector head, that is attached to an external electronics box.

# 4. X-RAY SPECTROMETER CHARACTERIZATION TECHNIQUES

## 4.1 Energy Resolution and Gain - $^{55}$Fe Source

Energy resolution near 5.9 keV and energy gain were confirmed from $^{55}$Fe measurements at LASP in Boulder, CO, USA. Figure 4 shows a representative MinXSS measurement of $^{55}$Fe $K_\alpha$ and $K_\beta$ complexes measured near 5.9 and 6.4 keV. The spectral resolution was estimated by calculating the FWHM of each line, assuming a Gaussian profile, and our measured values are consistent with Amptek's specifications for our X123s with their respective peaking times. The fitted energy gain yields an average value near ~0.029 keV/bin which is consistent with the expected 0.03 keV/bin. The offset value appears to be dependent on electrical grounding conditions and thus one should not expect the same energy offset for all measurement conditions (i.e., detector outside of the MinXSS spacecraft, inside the MinXSS spacecraft in the LASP Lab, SURF lab or on orbit). These results are encouraging and build confidence in our ability to separate the spectral signatures of certain line groups in the solar soft X-ray flux.

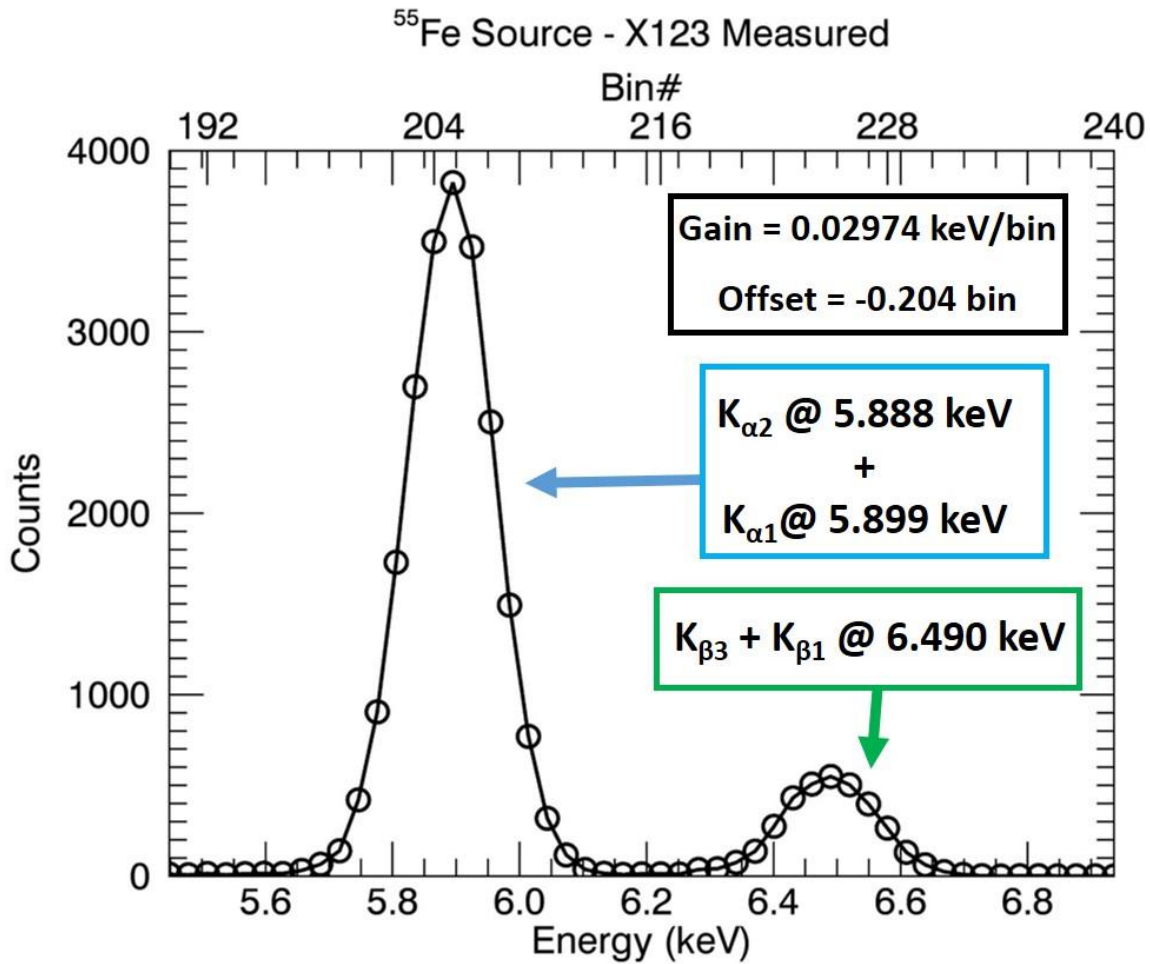

Figure 4. An example of an X123-measured $^{55}$Fe source emission profile. The $K_\alpha$ and $K_\beta$ line complexes at (~5.9 and ~ 6.5 keV respectively) are fully resolved from one another, but the individual components of these lines complexes are not, at the ~0.15 keV FWHM nominal spectral resolution at 5.9 keV. The fitted energy gain of ~0.029 keV/bin is consistent with the nominal binning of 0.03 keV bin.

## 4.2 NIST SURF

NIST SURF is a highly accurate and precise ultraviolet photon source[9] known to within 2%. and with extension in to the visible and soft X-ray ranges (soft X-ray accuracy known to within 10%). The synchrotron radiation produced by the relativistic electrons bent by magnetic fields is beamed into the direction of motion and directed down a beamline for instrument characterization tests. Basic descriptions of synchrotron radiation are given in Beskal et al. 2006[10], Attwood 2007[11] and references therein. The general proportionality relating the synchrotron beam current in milliamps, $I_e$, mean electron energy in mega-electron volts, $E_e$, and a photon spectral dependence function, F, involving modified Bessel functions, photon energy, $E_{ph}$, critical photon energy, $E_c$, and azimuthal (out of plane vertical angle), $\Psi$, to the number of generated photons, $N_{photons}$, in units of photons/s/mrad$_\theta$/mrad$_\Psi$, is given in Equation 1.

$$N_{photons} \propto I_e(mA) * E_e^2(MeV) * F(E_{ph}/E_c, \Psi) \quad (1)$$

Increasing the mean electron energy, $E_e$, changes the spectral distribution of the radiation (increasing $E_e$ produces more photons at every photon energy, but proportionally many more at high photon energies than at low photon energies). It must be noted that the mean electron energy, $E_e$, appears in the critical photon energy, $E_c$, also. This leads to the discussion of our first SURF test to determine our X123 spectral response, Multienergy.

### 4.2.1 Spectral Response Function – Multienergy Test

Initially we do not know the spectral response of our X123 spectrometer, but we can determine it by illuminating our detector with known spectral flux and dividing the measured counts by this known input flux. This is the ideal process for determining the instrument spectral response but is not always feasible due to low count rates leading to increasingly large uncertainties in the determined response from this "inversion" technique. Additionally, energy loss processes in the detector will yield counts at energies different than the initial photon energy (sometimes referred to as an off-diagonal response), which are not properly treated by a simple division. A more robust methodology is to use a model containing estimations of the dominant physical processes that influence the measured count flux and fit the best parameter values of the physical model to the measured count flux. We follow this ideology to estimate our X123 spectral response for many different input photon spectral distributions and respective measured count fluxes, which we call Multienergy.

Figure 5. Demonstrates the basic Multienergy procedure and the main steps are listed below.

1. Shine a known photon spectral intensity on the X123 aperture at a certain SURF mean electron energy.

2. Obtain the best fit parameters for the instrument response for that specific input photon flux that best reproduces the measured count flux.

3. Perform a Monte Carlo simulation around the best fit parameters to obtain 10 other estimations of the best fit parameters. This helps determine the uncertainty of the best fit parameters.

4. Repeat steps 1 – 3 for the next SURF mean electron energy.

5. Compute final best fit parameter values by taking the mean of the main fit parameters per energy (result of 2).

6. Compute the 'final' uncertainty on each best fit parameter by combining the uncertainty estimations from 2 (standard deviation of all energy best fit values) and 3 (MC estimated uncertainty) in quadrature (this could provide a conservative, overestimate of our uncertainty).

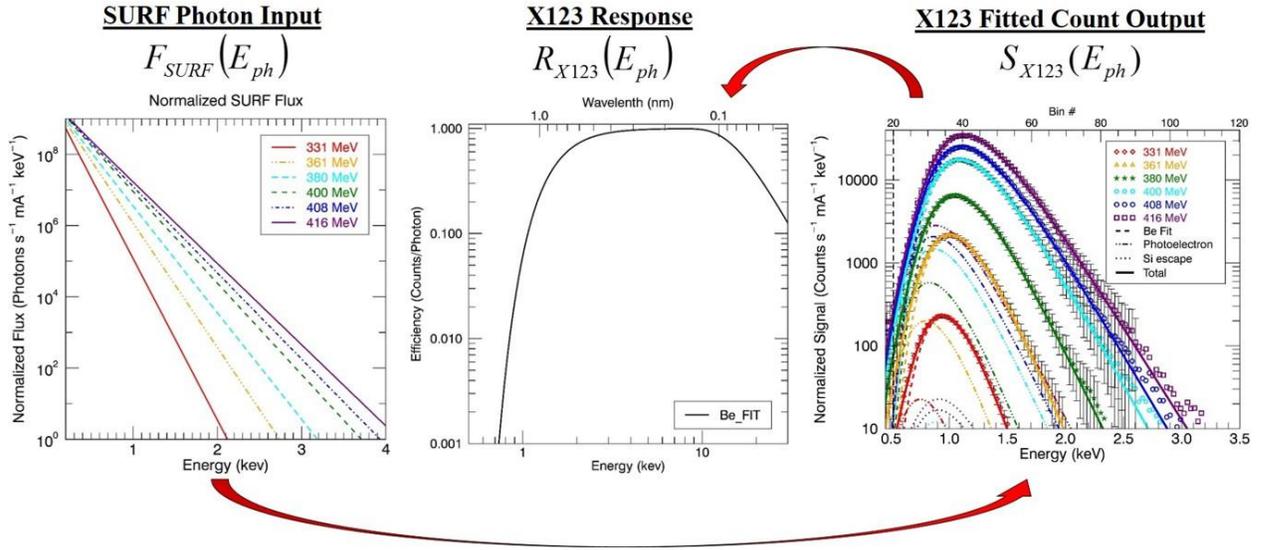

Figure 5. Outline of NIST SURF Multienergy technique to estimate the X123 spectral response. We fit a model of detector response parameters to the measured count flux from the SURF photon flux. Fitting the measured counts over the 0.5 - ~3.0 keV spectral range allows us to estimate the entire spectral efficiency from 0.5 – 30 keV from the model parameters. The main fit parameter is the Be window thickness. Other subsidiary components include contributions from escape of Si fluorescence photons from K and L (2s and 2p) excitations and photoelectrons produced in the Be window. The specifics of our model components will be discussed in a future paper.

The MinXSS X123 modeled count flux is determined from Equation II, where the measured X123 count flux from SURF, $S_{X123}$, in units of counts/s/cm$^2$/keV$_{bin}$, depends on the input SURF photon flux, $F_{SURF}$, units of photons/s/cm$^2$/keV, the X123 aperture geometric area, $A_{X123}$, in units of cm$^2$, the X123 response function $R_{X123}$, in units of counts/photon, and the X123 energy bins, $\Delta E_{bin}$, units of keV. Best fit values are obtained by minimizing the residuals between the beam-current-normalized measured count flux and the model-estimated count flux, forward modeled from the beam-current-normalized input SURF flux convolved with the modeled instrument response.

$$S_{X123}(E_{ph}) = \sum_{E_{ph}}[F_{SURF}(E_{ph}) * I_e(mA) * A_{X123} * R_{X123}(E_{ph})] * \Delta E_{bin} \qquad (2)$$

Table 1 lists the results of the Multienergy fit for the Be window thickness. The Be window thickness is the strongest dependent variable for the X123 spectral response. We compute the mass attenuation coefficient per photon energy following the analysis provided by Henke et al. 1993[12] and proceed to fit the Be window thickness from transmission and absorption properties. The detector bulk silicon and silicon oxide transmission and absorption properties are also taken from Henke et al. 1993 calculations and are fixed at the Amptek stated nominal thickness (500 μm for the depletion depth of the SDD). Any uncertainties in the silicon absorption parameters would affect the response at the high energies (≳10 keV), which we cannot measure with SURF.

Other subsidiary components of the instrument response function, $R_{X123}$, include Si escape processes of fluorescence photons and an estimated photoelectron contribution from the Be window. The Si-K shell (1s) and Si-L shell (2s and 2p) photon escape processes are incorporated by computing the relative bulk mass (volume averaged rate) fluorescence photon escape rate for the active volume of the detector relative to the Si photopeak (all of the incident photon energy being absorbed in the Si active area and detected by the readout mechanism) efficiency. Thus, the Si escape processes' contribution changes dynamically for each iteration of the detector response value during the fitting process. The probability for the Si escape process (mass attenuation coefficient contributions) are derived from Yeh and Lindau 1985[13] and Henke et al. 1993[12] calculations.

Methods of repelling possible photoelectrons generated in the Be window by an additional electric grid have not been implemented in the MinXSS CubeSat architecture. Thus, it is possible that photoelectrons generated in the Be window can eventually make it to the surface of the Be window, on the detector side, escape the surface and interact with the bulk Si if the electrons are energetic enough to overcome the negative detector surface bias voltage, the dead Si layer, and deposit their energy in the active region of the SDD. We dynamically estimate this contribution using a simple model coupled to the fitted Be window thickness. The Spicer model, Krowlikoski and Spicer 1969[14], is used to evaluate the probability of photoelectrons being created and eventually depositing their energy in the SDD active volume. Electron-electron scattering is taken as the dominant interaction process to inhibit the migration of electrons in the respective media with the mean free path calculated from asymptotic fits from Zjaja et al. 2006[15].

Table 1. Beryllium (Be) thickness fit results from the Multienergy SURF flux input to the X123. 10 Monte Carlo (MC) trial analysis was performed about each beam energy best fit to improve estimates of the Be fit thickness uncertainty. The Final Estimate listed of the best fit Be thickness is the mean of the best fit values from 361 – 416 MeV. The Final Uncertainty is the standard deviation of the single energy fits added in quadrature with the propagated uncertainty (derived uncertainty of the averaged best fit MC fits) from the MC trials. * = signifies that all energies except 331 MeV where used to construct the final value.

| SURF Multienergy Results | | |
|---|---|---|
| Beam Energy (MeV) | Best fit Be thickness (µm) | MC Best fit thickness (µm) |
| 331 | 24.62 | 25.09 |
| 361 | 24.74 | 24.42 |
| 380 | 25.08 | 25.18 |
| 400 | 25.34 | 25.50 |
| 408 | 25.35 | 25.53 |
| 416 | 25.48 | 25.43 |
| Average estimate* | 25.2 | 25.21 |
| Uncertainty of energy fit type* | 0.3 | 0.56 |
| Final Estimate* | 25.2 | |
| Final Uncertainty* | 0.64 | |

### 4.2.2 Detector Linearity Test

Detector measured counts are desired to scale linearly with the input photon flux. Equation I shows that the SURF flux scales linearly with the beam energy, thus we can test the linearity of the X123 response by increasing input SURF photon flux. Figure 6 shows the X123 fast and slow counter responses to an increasing photon flux. The fast counter has a peaking time of 100 ns and a pulse pair resolving time of ~120 ns, so photons that arrive more than 120 ns apart can be easily resolved as two different events by the fast counter. The slow counter peaking time for the data in Figure 6 is 4.8 µs and thus has much better spectral resolution properties than the fast counter, but at the cost of slower shaping. The slow counter data is used to make the X123 spectrum, and multiple photons arriving within the ~4.8 µs slow shaping time, detected as multiple events in the fast channel, are automatically rejected from the spectrum. Thus, if the slow counter counts diverge (negatively) from the fast counter counts for high count rates, then pulse pileup is degrading the quality of the resultant spectra. Additionally, for extremely large count rates, X123's slow channel dead time increases and becomes parasitic. Thus, it is important to diagnose our critical count rate. The X123 critical count rate is where pileup becomes significant (two or more photons are absorbed in the detector active area at a rate faster than the slow counter can resolve the energy content of the initial photon detection) and should be avoided in normal operations.

The X123 critical count rate is shown by the red horizontal line in Figure 6, and is around 8,000 counts/s. Prior solar measurements from a different X123 detector, with a thinner Be window thickness and larger aperture, on two NASA sounding rocket flights from Caspi et al. 2015[16], showed 1,100 counts/s measured for a GOES B1.7 level and 3,700 counts/s for a GOES B7 level. Thus we should not have count rate issues until GOES M levels.

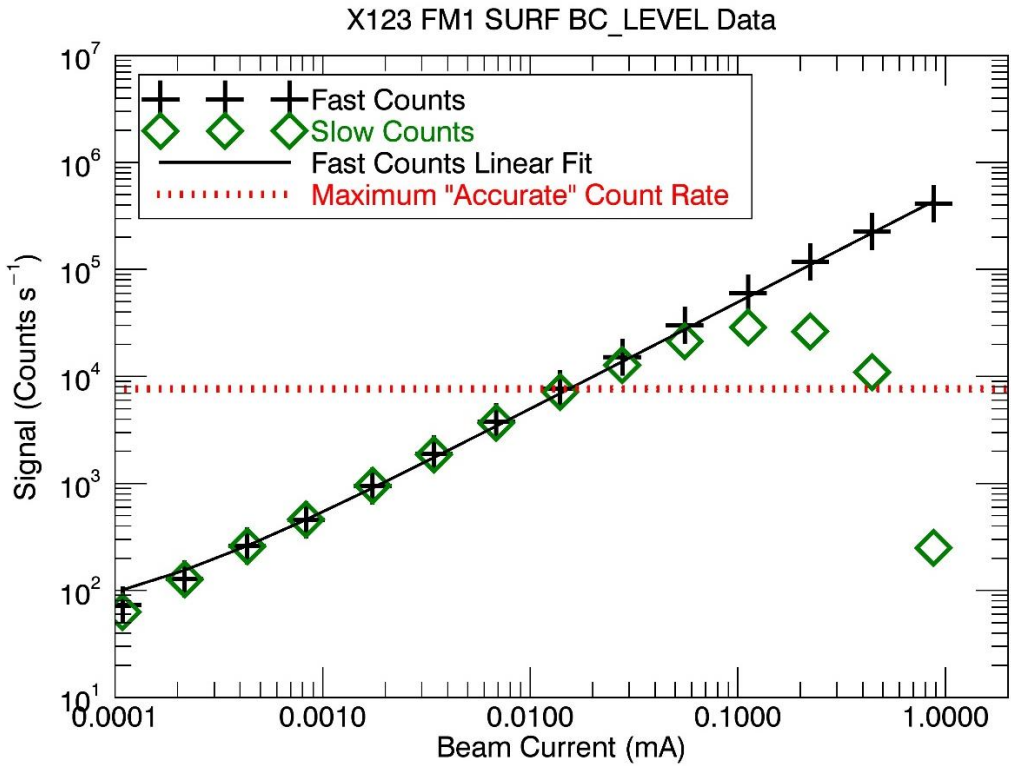

Figure 6. NIST SURF X123 linearity test shows a maximum count rate limit of ~8,000 counts/s, for an accurate spectrum. Maximum count rate is deduced from the last measured count rate, the red dotted lines, where the X123 slow counter (peaking time = 4.8 µs), the green diamonds, and fast counter (peaking time = 100 ns), the black pluses, begin to diverge. The black solid line is a linear fit to the fast counts. The SURF flux spectrum depends linearly with the SURF beam current (I), so a linear increase in the SURF beam current leads to a linear increase in the SURF photon flux.

## 5. PHOTON FLUX ESTIMATION FROM MEASURED COUNTS

Accurate determination of the MinXSS X123 spectral response gives us the ability to estimate the photon flux incident on the X123 aperture from the *measured* X123 count flux. This inversion technique is independent of forward fitting a spectral model to the measured count flux, where many spectral models are dependent on assumed plasma parameters (temperature, densities, chemical abundances, etc.), the spectral lines included in the model and experimentally determined atomic parameters (like oscillator strengths). Thus, MinXSS measurements can be used as an irradiance product and to help improve spectral models such as the CHIANTI Atomic Spectral Database[17]. Figure 7 is an example of the ability to invert the measured count flux to estimate the photon flux from the SURF data. Inverted photon flux estimates are created from the individual X123 response function components listed in Section 4.2.1. This demonstrates that we have confidence in estimating the input photon flux down to 1.0 keV.

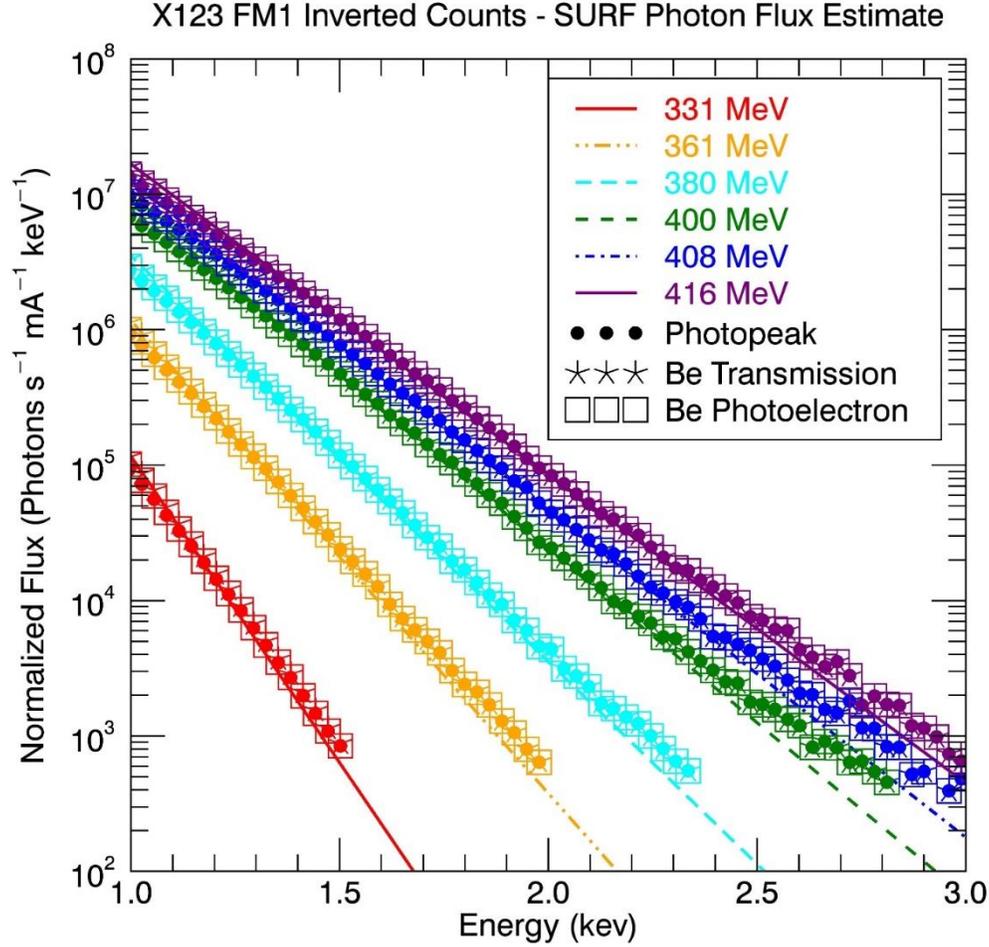

Figure 7. Estimation of the SURF flux incident on the X123 aperture, inverted from measured counts. The solid lines are the known SURF input flux, the symbols are three different response model aspects used for the inversion. The photopeak component (corrected for Si escape effects and Be photoelectrons) are the filled circles, the Be transmitted photon flux component are the asterisks and the Be photoelectron component are the open squares. All three model component inversions agree well down to 1 keV. Thus, we can confidently invert to estimate the photon flux over these energies.

## 6. MINXSS-2 AND THE FUTURE

As of June 2016, MinXSS-1 is in normal science operation mode and conducting solar soft X-ray measurements. MinXSS-2 is a twin version of MinXSS-1, but includes an improved X123 FAST SDD detector from Amptek and improved flight software from lessons learned from MinXSS-1. MinXSS-2 is scheduled to launch near the end of 2016 for a longer mission of 4 – 5 years in a sun-synchronous orbit. With the "faster" version of the X123, count rates of about four times higher can be accurately measured at a similar spectral resolution of the MinXSS-1 X123. The combined observing time from the MinXSS-1 and MinXSS-2 missions could result in up to 6 years of near continual spectrally resolved soft X-ray measurements to observe numerous possible spectral dependencies on the minimum and rising phase of the next solar cycle. MinXSS data can help demonstrate the need for future soft X-ray mission observatories. Additionally, the MinXSS CubeSat measurements can illustrate the importance of continual spectrally resolved soft X-ray measurements over many decades to build a data catalogue. This would necessitate newer, improved smaller missions to create regular continual spectrally resolved soft X-ray measurements to complement and be operated similar to the GOES XRS measurements.

## ACKNOWLEDGEMENTS

We would like to thank the many students and staff at CU Boulder and the NIST SURF team for their support in building and testing the MinXSS CubeSat instruments. Support for C. S. Moore was provided through NASA Space Technology Research Fellowship (NSTRF) Program Grant # NNX13AL35H, A. Caspi was also partially supported by NASA grant NNX15AK26G and support for the other authors is provided through NASA MinXSS grant # NNX14AN84G.